\documentstyle[prl,twocolumn,aps]{revtex}
\input{epsf}
\begin{document}
\draft
\twocolumn[\hsize\textwidth\columnwidth\hsize\csname @twocolumnfalse\endcsname
\title{Efficient quantum computation of high harmonics of the Liouville 
density distribution}

\author{B. Georgeot and D. L. Shepelyansky}

\address {Laboratoire de Physique Quantique, UMR 5626 du CNRS, 
Universit\'e Paul Sabatier, F-31062 Toulouse Cedex 4, France}

\date{October 16, 2001}

\maketitle

\begin{abstract}
We show explicitly 
that high harmonics of the classical Liouville density distribution
in the chaotic r\'egime can be obtained efficiently on a quantum computer
\cite{us,moriond}.
As was stated in \cite{us}, 
this provides information unaccessible for classical computer simulations,
and replies to the questions raised in \cite{diosi,zalka}.
\end{abstract}
\vskip1pc]

\narrowtext

In our Letter \cite{us} we showed on the example of Arnold cat map that
classical chaotic dynamics of exponentially many orbits can be simulated 
in polynomial time on a quantum computer.  
The Liouville density distribution $P(x_i,y_j)$ is encoded on a 
discretized lattice ($2^{n_q} \times 2^{n_q}$) 
using $3 n_q -1$ qubits organized
in three registers.   After each map iteration, 
the distribution is coded  in the
quantum state $\sum_{i,j} a_{ij} |x_i> |y_j>|0>$ with $P(x_i,y_j)=a_{ij}$ and 
$x_i=i/N$, $y_j=j/N$, $N=2^{n_q}$ \cite{note}.
One measurement of qubits in this basis gives one point in the phase
space and therefore the distribution $P(x_i,y_j)$ 
can be obtained approximately 
in polynomial number of measurements.  However the same information can be 
obtained via classical Monte Carlo simulation with 
a polynomial number of orbits,
as it was discussed by us in \cite{moriond} and later repeated in  
\cite{diosi}.  Based on this observation, the comment \cite{diosi}
makes a general claim that no new
information can be extracted efficiently from quantum computation of such 
classical maps ({\em paragraph 3 in} \cite{diosi}) (see also the comment
\cite{zalka}).  Here we show that 
this statement is incorrect.
Indeed, the quantum Fourier transform (QFT) of $a_{ij}$ provides
{\em nondiagonal 
observables} \cite{us}, namely the Fourier components 
$\tilde{P}(k_x,k_y)=\sum_{i,j} \exp(i2\pi (k_x x_i +k_y y_j))a_{ij}/N$. 
They obviously
contain important information relevant for the classical dynamics, and require
$O(n_q^2)$ operations including measurement.  
On the contrary,
all known classical algorithms will require exponential number of operations
 to obtain correct probabilities at high harmonics $k_{x,y} \sim N$.   
Such harmonics are important since due to chaos a significant part of total
 probability is transfered to
wavevectors with $k \sim \exp (h t)$ where $h$ is 
the Kolmogorov-Sinai entropy, 
and $t$ is the number of iterations (see  \cite{lieberman}).   
We note that the claim of
\cite{diosi} applies equally to the Shor algorithm, where all information is
also encoded only in squared moduli of amplitudes, but where the
QFT produces classically unaccessible information.

In Fig.1 we present the probability $\tilde{P}(k_x,k_y)$ in Fourier space for
different times $t$.  We note that  a 2-dimensional (2d) 
QFT can be efficiently 
implemented by application of usual QFT to each register consecutively.
The results show that  $\tilde{P}$ is composed of well-pronounced peaks,
most of which move with time to high wavevectors $k$.  They remain stable
in presence of noise in the quantum gates (e.g.  top right panel in Fig.1
is unchanged if $1\%$ noise is added in each gate).  The location and amplitude
of these peaks  can be extracted from a polynomial number of measurements
of qubits after the 2d QFT.

\begin{figure}
\epsfxsize=2.86in
\epsfysize=4.3in
\epsffile{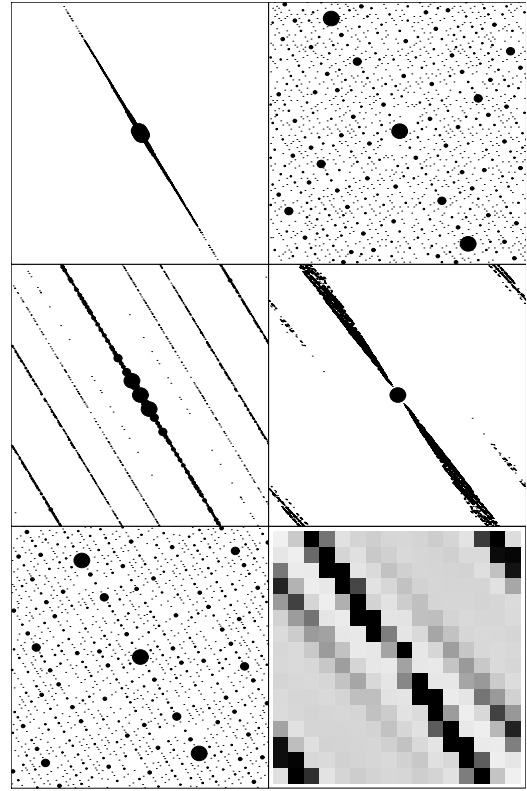}
\vglue -0.1cm
\caption{Fourier coefficients $|\tilde{P}(k_x,k_y)|^2$ 
of Liouville distribution
for $-N/2\leq k_{x,y} \leq N/2$, initial state as in Fig.1 of $[1]$.
Left column: cat map at $t=3,5,7$ from top to bottom for $n_q=10$.  Top right:
same at $t=5$, $n_q=7$. Middle right: $|\tilde{P}(k_x,k_y)|^2$ for 
perturbed cat map (see text) at $t=5$, $n_q=10$.  Peaks are shown by circles;
maximal circle size marks peaks with $1>|\tilde{P}(k_x,k_y)|^2>0.1$, 
circles twice smaller those with $0.1>|\tilde{P}(k_x,k_y)|^2>0.01$, etc...
Bottom right: coarse-grained image of $|\tilde{P}(k_x,k_y)|^2$ (proportional to
grayness) for the 
data of middle right panel, $n_f=4$. 
}
\label{fig1}
\end{figure}

For the Arnold cat map the dynamics in $(k_x,k_y)$ space is especially simple, 
given by $\bar{k_x}=k_x-k_y , \;
\bar{k_y}=2k_y-k_x \;\mbox{(mod} 
\;\mbox{N)}$.  However, generally this dynamics  is very complicated.  To 
exemplify this, we simulated the perturbed cat map 
$\bar{y}=y+x+x^2 , \;\bar{x}=x+\bar{y} \;\mbox{(mod} 
\;\mbox{1)}$ (Fig. 1).  
It can be iterated in $O(n_q^2)$ operations 
on a quantum computer using modular multiplication.  In this case,
main peaks can be seen directly for short times, while for larger times
a polynomial number of measurements of the first $n_f$ qubits \cite{moriond}
 gives a coarse-grained 
image of $|\tilde{P}(k_x,k_y)|^2$, including very high harmonics, 
which are unaccessible to classical computation \cite{note2}.  

As concerns the issue of errors in the computation, raised in \cite{zalka},
it should be stressed that the main aim of \cite{us} was to compare
the errors natural for classical and quantum computers.  In fact,
we showed that the natural/minimal (last bit) errors for classical
computer grow exponentially with number of map iterations
while the natural errors in quantum gates do not destroy
the time reversibility (Fig. 1 of \cite{us}).
As a further example, Fig. 1 of \cite{demon} 
clearly shows that the errors of relative precision
$10^{-8}$ (comparable with ordinary precision on Pentium III) completely
destroy the reversibility of classical dynamics.  At the same time,
the quantum errors of relative precision $10^{-2}$ in operations
of quantum gates preserve the reversibility.
Of course, since the quantum algorithm simulates the classical
discretized map {\em exactly}, the last bit errors made on a quantum 
computer lead to exponential divergence of nearby trajectories
and exponential drop of fidelity.  This is clearly illustrated
by Fig. 3 in \cite{us}.  Thus exponential instability of classical
chaos is preserved in quantum simulations.  
The quantum computer has enormous capacity in memory and
precision which grows exponentially with the number of qubits;
thus in the quantum case the 
last bit errors are much less important than for the classical
computer with its limited memory space.
However, the quantum computer
has its own natural errors related to imperfections 
in gate operations.
The results presented in \cite{us,moriond,demon} show that the
 quantum simulation is stable with respect to its natural errors
in quantum gates operations while the classical computation is unstable
with respect to its natural errors in the last bit of dynamical variables.

This work was supported in part by the NSA
and ARDA under ARO contract number DAAD19-01-1-0553.


\begin{thebibliography}{99}
\bibitem{note}An initial distribution with exponentially many orbits
can be prepared efficiently on a quantum computer; for example a square
of size $N/2\times N/2$ can be built 
by using $2n_q -2$ Hadamard gates acting on the last $n_q -1 $ qubits of each
register $x,y$.
\bibitem{note2}Contrary to what was argued in  \cite{zalka}, high harmonics
cannot be found from the fine structure of ``a few closely spaced
final points'', since the Fourier coefficients involve the full distribution.
\end{thebibliography}
\end{document}